\documentclass[10pt,aps,prd,twocolumn,showpacs,superscriptaddress,nofootinbib,nobibnotes,longbibliography,floatfix]{revtex4-1}

\usepackage[utf8]{inputenc}
\usepackage[T1]{fontenc}
\usepackage{comment}
\usepackage{bm}
\usepackage{mathtools,amsmath,amssymb,amsfonts,mathrsfs,eucal,graphicx,tensor,csquotes,accents,commath,chngcntr,siunitx}
\usepackage[dvipsnames]{xcolor}
\usepackage[unicode]{hyperref}
\hypersetup{colorlinks=true, citecolor=MidnightBlue,
            linkcolor=MidnightBlue, urlcolor=MidnightBlue, linktocpage=true}
\usepackage[normalem]{ulem}

\DeclareMathAlphabet{\mathpzc}{OT1}{pzc}{m}{it}
\definecolor{darkgreen}{rgb}{0.0, 0.6, 0.0}

\begin{document}
\title{Hairy Black Holes: Non-existence of Short Hairs and Bound on Light Ring Size}

\author{Rajes Ghosh}
\email{rajes.ghosh@iitgn.ac.in}
\affiliation{Indian Institute of Technology, Gandhinagar, Gujarat 382055, India.}

\author{Selim Sk}
\email{selimsk@iitgn.ac.in }
\affiliation{Indian Institute of Technology, Gandhinagar, Gujarat 382055, India.}

\author{Sudipta Sarkar}
\email{sudiptas@iitgn.ac.in}
\affiliation{Indian Institute of Technology, Gandhinagar, Gujarat 382055, India.}

\begin{abstract}
Several hairy black hole solutions are known to violate the original version of the celebrated no-hair conjecture. This prompted the development of a new theorem that establishes a universal lower bound on the extension of hairs outside any $4$-dimensional black hole solutions of general relativity. Our work presents a novel generalization of this ``no-short hair'' theorem, which notably does not use gravitational field equations and is valid for arbitrary spacetime dimensions ($D \geq 4$). Consequently, irrespective of the underlying theory of gravity, the ``hairosphere'' must extend to the innermost light ring of the black hole spacetime. Various possible observational implications of this intriguing theorem are discussed, and other useful consequences are explored.

\end{abstract}
\maketitle

\section{Introduction}
Einstein's theory of general relativity (GR) is the cornerstone of our current understanding of classical gravitational physics. One of the most striking predictions of GR is the existence of black holes (BHs), which exhibit remarkable simplicity in stark contrast to other relativistic configurations. For example, in four dimensions, the stationary and asymptotically flat spacetime outside a vacuum BH solution of GR is represented by a Kerr metric characterized by only two parameters, namely the mass $M$ and spin $a$. As a consequence, all higher multipoles of Kerr BHs are uniquely determined by $M$ and $a$ \cite{Carter:1971zc, Robinson:1975bv, Geroch:1970cd, Hansen:1974zz}.\\

Even in the presence of matter, this important result finds its extension in the form of a well-motivated belief, known as the no-hair hypothesis \cite{Ruffini:1971bza, Carter:1973rla} (see also \cite{2010GReGr..42..653C}): Irrespective of the nature of the matter content, the end product of a gravitational collapse within the framework of GR can be completely specified by conserved charges such as mass, angular momentum, and electric charge measured at asymptotic infinity without any additional parameters (referred to as ``hairs''). The motivation behind this conjecture stemmed from previous uniqueness theorems concerning BH solutions within the framework of GR \cite{Israel:1967wq, Robinson:1974nf}. Additionally, a heuristic physical interpretation was suggested in Ref.~\cite{Nunez:1996xv} that any matter fields existing outside a newly formed BH would either be emitted away to spatial infinity over time or absorbed by the BH itself unless those fields were associated with conserved charges at asymptotic infinity. The initial support for the no-hair conjecture came from the work of Bekenstein \cite{Bekenstein:1971hc, Bekenstein:1972ky}, which states that a stationary BH can not be endowed with any exterior scalar, vector, or spin-$2$ meson fields. These results led to a belief that the no-hair conjecture is true irrespective of the nature of the matter content.\\

The first counterexample to this conjecture was provided by the discovery of the ``colored'' BH solution in Einstein-Yang-Mills theory, which contains an additional integer parameter that is not associated with any conserved charges like mass and spin \cite{Bizon:1990sr}. It was soon established that this original version of the no-hair conjecture is invalid, and there are several hairy BH solutions, which include BHs with skyrmion \cite{Luckock:1986tr}, dilaton hairs \cite{Kanti:1995vq}, and axion \cite{Campbell:1991rz}.\\

Moreover, beyond the framework of GR, the presence of any putative modifications might lead to the violation of the no-hair property, resulting BHs with extra hairs. Over the years, such hairy modifications of Kerr BHs have been extensively studied, and their observational signatures have been searched for \cite{Ryan:1995wh, Ryan:1997hg, Gair:2007kr, Bambi:2011jq, Bambi:2011vc, Bambi:2015ldr, Isi:2019aib, EventHorizonTelescope:2021dqv}. However, these distinctive non-Kerr features arising from the presence of extra hairs can be suitably captured in observations probing only the far-away regions of spacetime, if the hairs extend sufficiently outside the horizon. Thus, the essential question of observational relevance is as follows: Can BHs have short hair confined only in the near-horizon region? An affirmative answer to this question would essentially imply that BHs with short hairs mimic Kerr-like signatures when probed in the far-field regions, though the near-horizon structure could be very different from that of a Kerr BH. Given this crucial observational relevance, we want to investigate this question in detail.\\ 

Given the violation of the no-hair conjecture, it is only natural to search for the crucial physical attribute which led to the existence of these solutions. In this regard, an important observation made in Ref.~\cite{Nunez:1996xv} is that the non-linear character of the matter fields plays a fundamental role in the construction of the hairy black-hole solutions. There exists a self-interaction between the part of the field very close to the horizon and the part of the field relatively far away. This interaction prevents the near-horizon hair from collapsing into the BH while ensuring that the far-region hair does not escape to infinity. Using the Einstein field equations, the weak energy condition (WEC) and the non-positive trace condition on matter, it was shown that BHs cannot have short hairs and the region (referred to as ``hairosphere'') having the non-linear behavior must extend at least up to three-halves of the horizon radius \cite{Nunez:1996xv}, which curiously corresponds to the location of the light ring (LR) of a vacuum Schwarzschild BH in $D=4$ dimensions. However, this curious fact is not just a mere coincidence, as it was later proved explicitly that the hairosphere must always extend beyond the innermost LR, for any static non-vacuum (spherically symmetric) hairy BH configurations of GR in $D=4$ dimensions \cite{Hod:2011aa}.\\

This ``no-short hair'' theorem in GR provides a lower bound to the extent of the hairosphere outside the BH horizon. Thus, to detect the presence of hair around BHs, it is sufficient to probe till the near-LR region alone, which is greatly exemplified by the BH shadow observations. In other words, since the hairy configuration is extended at least up to the LR, it is possible to probe the presence of the hair in the image of the BHs \cite{Cunha:2015yba,  Cunha:2019ikd,  EventHorizonTelescope:2021dqv}.\\

Interestingly, with the advent of unprecedented technological progress, the question stated above becomes very relevant for observational purposes as we are gradually probing the near-field regime of BHs with ever-increasing precision. The detection of gravitational waves by the LIGO-Virgo collaborations \cite{LIGOScientific:2016aoc} and the imaging of BH shadow by the Event Horizon Telescope (EHT) \cite{EventHorizonTelescope:2019dse} have opened up a new era in BH physics. These observations may allow us to test the no (short) hair conjecture and find the observational signature of hairy BHs.\\

Motivated by both the theoretical and observational importance of the aforesaid results in GR \cite{Nunez:1996xv, Hod:2011aa}, we would like to ask the following question: Do the field equations or dimensionality of the spacetime play any fundamental role in determining the length of the hair being short (confined solely to near-horizon regime) or long (extends ``sufficiently'' beyond the horizon)? In particular, we want to know whether a similar result to no-short hair theorem can be obtained \textit{independent} of the gravitational field equations in any dimensions ($D \geq 4$). If true, this will provide a novel generalization of the theorem to any theory of gravity which admits hairy BH solutions.\\

Intriguingly, the answer to the above question is affirmative. We show that for any static, spherically symmetric and asymptotically flat hairy BH in $D$-dimensions, regardless of the theory of gravity under consideration, the hairosphere must extend at least up to the innermost LR. Therefore, exploring the near-LR regions alone may give us useful information about the signatures of BH hairs in any theory of gravity (similar to GR).  Such an extension is important on two fronts. First, it helps us understand the features of BH solutions of modified gravity in a unified theory-agnostic way. Second, such a generalization shows that the no-short hair result cannot be conceived as a null test of GR.\\

Other possible consequences of our analysis, such as existence of short hairs on horizonless compact objects is outlined generalizing the work of Ref.~\cite{Peng:2020hkz} in a theory-agnostic way, and various generalizations of the works of Ref.~\cite{Hod:2013jhd, Chakraborty:2021dmu} about size of LRs in several higher-dimensional/modified theories of gravity are explored.

\section{Hairy Black holes}
 A general static, spherically symmetric BH spacetime in $D$-dimensions is described by the metric

\begin{align} \label{Metric}
    ds^2\,=\, -\, f(r)\, dt^2\,+\,\frac{1}{k(r)}\,dr^2\,+\,h(r)\,d\Omega^2_{D-2}\, ,
\end{align}
where $h(r)$ is assumed to be an increasing function of the Schwarzschild-like radial coordinate $r$ everywhere outside the outermost (regular) horizon at $r\,=\,r_H$ such that $k(r_H) = 0$. Then, staticity implies that the norm of the timelike Killing vector must vanish at $r\,=\,r_H$, which implies $f(r_H)\,=\,0$ \cite{1968JMP.9.1319V}. We also assume that the spacetime is asymptotically flat, which requires that $f(r)\rightarrow 1$, $k(r)\rightarrow 1$, and $h(r) \sim r^2$ as  $ r \to \infty$.\\

The metric described by Eq.(\ref{Metric}) is a solution of a theory sourced by an arbitrary energy-momentum tensor $T_{\mu \nu}$. Due to the spherical symmetry, $T^\mu_\nu$ should  remain invariant under any rotation in the $(D-2)$-dimensional compact space with coordinates $\{ \theta_i\}, \, i = 1, \, 2, \, (D-2) $. Therefore, $T^t_{\theta_i}\,=\, T^r_{\theta_i}\,=\,0$, as they transform like vectors under such rotations. Also, spherical symmetry further implies $ T^{\theta_1}_{\theta_1}\,=\,T^{\theta_2}_{\theta_2}=\,\cdots=\,T^{\theta_{D-2}}_{\theta_{D-2}}$,  
and all the off-diagonal transverse components vanish. Thus, we have only four non-zero independent components of the $T^{\mu}_{\nu}$, namely $\{T^t_t :=-\rho\, ;\, T^t_r \, ;\, T^r_r := p\, ;\, T^{\theta_1}_{\theta_1 } := p_T \}$, where $\rho$, $p$, and $p_T$ are identified as the energy density, radial, and tangential pressure, respectively. All these components are functions of $r$ only. Now, the radial component of the energy-momentum conservation equation $\nabla_\mu\,T^\mu_\nu\,=\,0$, leads to

 \begin{align}\label{rconserve}
      \hat{P}'(r) \,=\,\frac{h^{D/2-1}}{2\, f}\,(p+\rho)\,\Delta\,+\, \frac{h^{D/2-1}}{2}\,h'\, T\, ,
 \end{align}
where $'$ represents radial derivative, $\hat{P}= h^{D/2}\, p$, $\Delta\,=\,(f\,h'\,-\,h\,f')$, and $T$ stands for the trace of the energy-momentum tensor. This equation generalizes the similar result of Ref.~\cite{Hod:2011aa} in the context of 4-dimensional GR to an arbitrary dimensions ($D \geq 4$) in a theory-independent way. That is, we have not used any gravitational field equations to derive Eq.(\ref{rconserve}).\\

We will assume that the matter content present in this BH spacetime satisfies the following conditions \cite{Nunez:1996xv, Hod:2011aa}:\\
\\
(i) The WEC implying the energy density to be positive semidefinite, $\rho \geq 0$. It also bounds the radial pressure via the inequality $p\,+\,\rho\geq 0$.\\

\noindent
(ii)  The trace of the energy-momentum tensor is non-positive, $T \leq 0$ implying $p + (D-2)\, p_T \leq \rho$. This assumption plays a crucial role in the existence of hair.\\

\noindent
(iii) The energy density $\rho$ approaches zero faster than $r^{-D}$ as $r\rightarrow\, \infty$. This condition naturally rules out the existence of any extra conserved charges \cite{Nunez:1996xv, Hod:2011aa}. Therefore, we have the boundary condition $\hat{P}(r) \rightarrow 0$ as $r\rightarrow \infty$. This also implies that we are working with a hairy BH solution.\\

In general, $\hat{P}(r)$ provides a useful tool to probe the region where the nonlinear behavior of the matter field is present. Therefore, it is essential to study the behavior of $\hat{P}(r)$ in the vicinity of the BH horizon. We define a local coordinate system near the horizon as $dx\,=\, k^{-1/2}\, dr$, known as proper radial distance. The Equivalence principle ensures that the proper radial distance is a well-defined coordinate. In terms of this new coordinate, Eq.(\ref{rconserve}) becomes

\begin{align}\label{radialp}
 \nonumber   \frac{d \hat{P}}{dx} \,=\,\frac{h^{D/2-1}}{2\,f}\,(p+\rho) \Big(f\,\frac{dh}{dx}\,-\,h\,\frac{df}{dx}\Big) \\+\, \frac{h^{D/2-1}}{2}\, T\, \Big(\frac{dh}{dx}\Big)\, .
 \end{align}

Our assumption of having a regular horizon at $r\,=\,r_H$ indicates physical invariants, such as $T_{\mu\nu}\,T^{\mu\nu}$, must be finite there. Therefore, all components of $T^\mu_\nu$, especially $\rho$ and $p$, should be finite at the horizon, implying L.H.S side of Eq.(\ref{radialp}) is also finite in the limit $r\rightarrow\,r_H$. Assuming a non-extremal BH with $f'(r_H)>0$, WEC and the finiteness of the R.H.S of Eq.(\ref{radialp}) at the horizon imply that
\begin{align}\label{rpH}
    p\,(r_H) = - \rho\,(r_H) \leq 0. 
\end{align}
The above equation along with the fact that $k(r \to r_H) > 0$ gives the following results,

\begin{align} \label{PH}
    \hat{P}(r \to r_H)\,\leq \,0,  \hspace{3mm}\textit{and} \hspace{3mm} \hat{P}'(r \to r_H) < 0, 
\end{align}
in the vicinity of the BH horizon. Using the conditions derived above, we now proceed to prove the key theorem of our paper.\\

\noindent
\emph{Theorem}: If the matter content satisfies all three conditions stated above and there exists a non-empty interval $r_H\,\leq\,r\,\leq\,r_p$ where the quantity $\Delta(r)\,=\,(f\,h'\,-\,h\,f')\leq \,0$, we must have $\hat{P}'(r_H\,\leq\,r\,\leq\,r_p)\leq\,0$.\\

The proof of this statement can be derived using Eq.(\ref{rconserve}) and Eq.(\ref{PH}). The WEC and the non-positivity of $\Delta(r)$ in the region $r_H\,\leq\,r\,\leq\,r_p$ imply the first term in the R.H.S of Eq.(\ref{rconserve}) is non-positive. The same is true for the second term as well due to the trace condition $T\leq 0$ and $h'(r)>0$. Then, it immediately follows that $\hat{P}'(r) \leq 0$ in the region $r_H\,\leq\,r\,\leq\,r_p$.\\

Though it remains to be shown that such a radial interval exists where $\Delta(r) \leq 0$, which we shall consider in subsequent paragraphs, let us first discuss its consequence. The above theorem along with Eq.(\ref{PH}) suggests that $\hat{P}(r)$ is non-positive at the horizon and then it decreases at least up to $r=r_p$. Now following Ref.~\cite{Nunez:1996xv, Hod:2011aa}, let us define the extend of the ``hairosphere'' $r_{\text{hair}}$ to be the radius at which $|\hat{P}(r)|$ has a local maximum, then we must have

\begin{align}\label{hair}
r_{\text{hair}}\,\geq\,r_p.  
\end{align}

Thus, the hair on an asymptotically flat, static, and spherically symmetric BH solution of any theory of gravity cannot be shorter than $r_p$. Thus, it is essential to know whether $r_p$ has any physical characteristic of the the BH spacetime. Though it is not apparent, we shall now show that $r_p$ corresponds to the location of the innermost LR.\\
 
To show this, we follow a similar analysis provided in Ref.~\cite{Hod:2011aa}, and study the timelike and null geodesics in the BH spacetime described by Eq.(\ref{Metric}). The motion of such a particle with energy $E$ and angular momentum $L$, moving in the equatorial plane is described by,
 \begin{align}
    \Dot{r}^2\,= \, k(r) \left[ \frac{E^2}{f(r)}\,- \, \frac{L^2}{h(r)}\,-\, \epsilon\right]\, ,
 \end{align}
where a dot denotes a derivative with respect to some affine parameter. Also, $\epsilon\,=\,0$ represents null geodesics and $\epsilon\,=\,1$ represents timelike geodesics. Circular geodesics are characterized by $\Dot{r}^2\,=0\,=\,(\Dot{r}^2)'$. Solving these two equations, we get

\begin{align}
    E^2\,=\,\frac{\epsilon\,f^2(r)\,h'(r)}{\Delta(r)}\, , \hspace{6mm}  L^2\,=\,\frac{\epsilon\,h^2(r)\,f'(r)}{\Delta(r)}\, .
\end{align}
Since for timelike/null orbits, $E^2$ and $L^2$ should be non-negative, we must have the quantity $\Delta(r)\,>\,0$ for timelike circular geodesic, and  $\Delta(r)\,=\,0$ for null circular geodesics. Note that, the zeros of the function $\Delta (r)$ denote the locations of the light rings (LR) of the spacetime. In order to analyse the behavior of $\Delta(r)$ in the region $[r_H, \infty)$, we consider two auxiliary functions: $L(r)\,=\,f(r)\,h'(r)$, and $R(r)\,=\,h(r)\,f'(r)$ such that $\Delta=L-R$. The behaviors of these functions are as follows,\\
\\
(i)  At the horizon $r\,=\,r_H$, $L(r_H)\,=\,0$, and $R(r_H)\,>\,0$.\\

\noindent
(ii)   For $r\,\rightarrow\,\infty$, $L(r)\,\sim\,r$, and $R(r)\,\sim\,r^{-(D-4)}$\, .\\

\noindent
The above conditions for $D\,\geq\,4$ suggest that $\Delta(r)$ has an odd number of zeroes, which correspond to the LRs outside the outermost horizon. These LRs divide the interval $[r_H, \infty)$ into an even number of regions. Since in the outermost region the quantity $\Delta(r)\,\geq \, 0$, we must have $\Delta(r)\,\leq \, 0$ in the innermost region $r_H \leq r \leq r_\gamma^1$, where $r_\gamma^1$ denotes the location of the innermost LR. This concludes our proof that there exists a radial interval $r_H \leq r \leq r_p$ with $\Delta(r)\,\leq\,0$, provided that we identify $r_p$ with $r_\gamma^1$. Therefore, the hairosphere of a BH must extend at least up to the inniermost LR, $r_{\text{hair}} \geq r_\gamma^1$. In other words, we have shown that BHs can not have short hairs confined only in the near-horizon region.\\

Some comments are as follows. Though our analysis runs parallel to the work of Ref.~\cite{Hod:2011aa}, the novelty of our work lies in its generality. Namely, we have not used any field equations to obtain the result and it is valid for any spacetime dimensions $D \geq 4$. Thus, our work provides a novel theory-agnostic generalizations of the no-short hair theorem of GR discussed in Ref.~\cite{Nunez:1996xv, Hod:2011aa}. Now, we shall discuss few interesting consequences of Eq.(\ref{rconserve}) in the following sub-sections.

\subsection{Size of Static Shells}
Consider a static shell of finite thickness located entirely between the outer horizon $r_H$ and the innermost LR $r^1_\gamma$, i.e., $r_H < r_1<r_2 < r^1_\gamma$, where $r_1$ and $r_2$ are the inner and the outer radius of the shell. Now, if the matter field obeys WEC and $T\,\leq\,0$ as discussed for the BH case, Eq.(\ref{rconserve}) implies to have $\hat{P}'(r_1 \leq r \leq r_2) \leq 0$. Consequently, this implies that the pressure at the outer surface is lower than (or same as) the pressure at the inner surface. However, this contradicts the requirement that the pressure must be zero at both surfaces, considering there is no matter present outside the shell. Therefore, we conclude that a static shell of finite thickness can not exist entirely between the outer horizon and the innermost LR. Note,  this result is obtained without using any field equations.

\subsection{Bound on LRs in EGB Gravity}

One can utilize our theorem in the context of any particular theories of gravity (thereby, using the corresponding field equations) to get an upper bound on the size of the innermost LR by performing a similar calculation presented in Refs.~\cite{Hod:2013jhd, Chakraborty:2021dmu}. For the purpose of illustration, let us consider the EGB theory given by $\mathcal{L}\,=\,R\,+\,\hat{\alpha}\,(R^2\,-\,4R_{\alpha\beta}\,R^{\alpha\beta}\,+\, R_{\alpha\beta\gamma\delta}\,R^{\alpha\beta\gamma\delta})$, which leads to the following field equation,
\begin{align}\label{EGB}
G^{(1)}_{\alpha\beta}\,+\,\hat{\alpha}\,G^{(2)}_{\alpha\beta}\,=\,8\pi\,T_{\alpha\beta}\, .
\end{align}
The explicit forms of the $G^{(1)}_{\alpha\beta}$ and $G^{(2)}_{\alpha\beta}$ can be found in Ref.~\cite{Zhou:2011wa}. As a solution of Eq.(\ref{EGB}), we consider a spherically symmetric, static and asymptotically flat BH metric in the form of

\begin{align} \label{EGBMetric}
    ds^2\,=\, -\, e^{-2\delta(r)}\,\mu(r)\, dt^2\,+\,\frac{1}{\mu(r)}\,dr^2\,+\,r^2\,d\Omega^2_{D-2}\, .
\end{align}
We assume that there exists a regular non-extremal event horizon at $r\,=\,r_H$, so that $\mu(r_H)\,=\,0$, $\mu'(r_H)\,>\,0$, and $\delta(r)$ and its radial derivative is finite there \cite{Hod:2013jhd}. Due to the asymptotic flatness, we must have $\mu(r) \to 1$ and $\delta(r) \to 0$ at infinity. Now, using the field Eq.(\ref{EGB}), we get
\begin{equation} \label{field1}
\begin{split}
    &\delta' =\,- \frac{8\, \pi\, r^3\, (p+\rho)}{(D-2)\, \mu\, \left[r^2 + 4\, \alpha \, (1-\mu)\right]}\, ,\\
    \\
    & \mu' = \frac{2\, r^3\, (D-3)}{r^2\,+\,4\, \alpha \, (1-\mu)} \Bigg[ \frac{1-\mu}{2\, r^2}+ \frac{\alpha\,(D-5)\, (1-\mu)^2}{(D-3)\, r^4} \\ 
    &\kern130pt -\frac{8\, \pi\, \rho}{(D-2)(D-3)}\Bigg]\, ,
\end{split}
\end{equation}
where $T^t_t\,=\,-\rho$, $T^r_r\,=\,p$ and $\alpha\,=\, (D-3)(D-4)\hat{\alpha}/2$. One can easily find the location of the innermost LR $r_\gamma^1$, which is given by the smallest positive root of the following equation
\begin{align}\label{lightring}
   2\, e^{-2\delta_\gamma}\, \mu_\gamma - r_\gamma^1\, \Big(e^{-2\delta}\, \mu\Big)'_{r_\gamma^1}\,=\,0\, . 
\end{align}
Here, $\mu_\gamma$ and  $\delta_\gamma$ are shorthands for $\mu(r_\gamma^1)$ and $\delta(r_\gamma^1)$, respectively. Then, using Eqs.(\ref{field1}, \ref{lightring}), we get an identity,

\begin{align}\label{lightring2}
\nonumber N(r_\gamma^1)\,:=\, (D-1)\mu_\gamma\,-\,(D-3)\,+\,\frac{8\, \alpha\, \mu_\gamma(1-\mu_\gamma)}{(r_\gamma^1)^2}\\-\,\frac{2\, \alpha\, (D-5)(1-\mu_\gamma)^2}{\left(r_\gamma^1\right)^2}\,=\,\frac{16\, \pi\, r^2_\gamma\, p(r_\gamma)}{D-2}\, .
\end{align}
Then, our theorem implies $p(r^1_\gamma)\leq\,0$, which leads to the following condition: 
\begin{align} \label{boundp}
    N(r^1_\gamma)\leq\,0.
\end{align}
Now, to get an upper bound on the size of the innermost LR, we require an explicit form for $\mu(r)$. It is useful to define a mass function
\begin{align}
    m(r)\,=\,\frac{r_H}{2}\,+\, \Omega_{D-2} \int_{r_H}^r \rho(x)\,x^{D-2}\, dx\ ,
\end{align}
where $\Omega_{D-2}\,=\,2\pi^{(D-1)/2}/\Gamma[(D-1)/2]$ is the surface element of unit $(D-2)$-sphere and we have chosen the boundary condition, $m(r_H)\,=\,r_H/2\,>\,0$. With these conditions, the $\mu'$-equation given by Eq. (\ref{field1}) admits the following solution,
\begin{align}\label{EGBf}
    \mu(r)\,=\,1\,+\,\frac{r^2}{4\, \alpha}\Bigg[ 1-\sqrt{1+\frac{16\, \alpha\, M(r)}{r^{D-1}}}\, \Bigg]\, ,
\end{align}
where $M(r)\,=\,8\pi m(r)/(D-2)\Omega_{D-2}$.
Replacing Eq.(\ref{EGBf}) in Eq.(\ref{boundp}), we get the following polynomial inequality

\begin{align}\label{poly}
    (r^1_\gamma)^{2D-6}\,+\,16\, \alpha\, M_\gamma\,  (r^1_\gamma)^{D-5} \,-\,(D-1)^2\, M^2_\gamma\,\leq\,0\, ,
\end{align}
where $M_\gamma = M(r_\gamma^1)$. The above equation has some important consequences. An immediate one would be the innermost LR of any D-dimensional BHs in GR ($\alpha\, =\, 0$) is bounded above by the radius ($r_\gamma^{ST}$) of the LR of Schwarzschild-Tangherlini (ST) BH \cite{Tangherlini:1963bw} with mass $M_{ST} := M(r \to \infty)$. It follows directly from the $\alpha \to 0$ limit of the above inequality that $r_{\gamma, (GR)}^1 \leq \left[(D-1)\, M_\gamma \right]^{1/(D-3)} = r_\gamma^{ST}$, as $M_{ST}\,\geq\, M_\gamma$. This generalizes the result of Ref.~\cite{Hod:2013jhd} for higher dimensional GR.\\

Now, for EGB gravity with $\alpha\, \neq\, 0$, we must choose $\alpha$ in such a way that there exists a horizon (non-extremal) to avoid naked singularity. In this context, it is possible to generalize the above result of GR and demonstrate that Boulware-Deser (BD) BH \cite{Boulware:1985wk} has the largest LR. For this purpose, let us consider the following polynomials in $r\,\in [0,\infty)$ as
\begin{align*}
    F_1(r)\,&=\,r^{2D-6}\,+\,16\, \alpha\, M_\gamma\, r^{D-5}\,-\,(D-1)^2\, M^2_\gamma\, , \\
    F_2(r)\,&=\,r^{2D-6}\,+\,16\, \alpha\, M_{BD}\, r^{D-5}\,-\,(D-1)^2\, M^2_{BD}\, , 
\end{align*}
where $M_{BD}\,:=\,M(r\to \infty)\,\geq M_\gamma$. For either signs of $\alpha$, Descartes' rule of signs ensures that both $F_1$ and $F_2$ have single positive root $r_m$ and $r_\gamma^{BD}$, respectively. Note that $r_\gamma^{BD}$ denotes the location of the LR of the BD BH with mass $M_{BD}$. Whereas Eq.(\ref{poly}) suggests that the innermost LR of any BH solution of EGB gravity is bounded above by $r_m$, i.e., $r_\gamma^{1} \leq r_m$.\\

Since $M_{BD} \geq M_\gamma$, the function $F_2(r=0)$ has a value less than $F_1(r=0)$. Then, it is suggestive to evaluate $F_2(r)$ at the location of the root ($r_m$) of $F_1(r)$. Some simple algebraic manipulation gives us, 

\begin{equation}
    F_2(r_m) = \left[\frac{r_m^{2D-6}}{M_\gamma} + (D-1)\,M_{BD}\right]\,\left(M_\gamma-M_{BD}\right)\, .  
\end{equation}
Since $M_{BD}\,\geq\,M_\gamma$, we have $F_2(r_m)\,\leq\,0$. This, in turn, implies the following inequality must holds: $r_\gamma^{BD} \geq r_m \geq r_\gamma^1$, which completes the proof that the size of the innermost LR of BH solutions of EGB gravity is bounded above by that of Boulware-Desser LR.\\

\subsection{Comments on Horizonless Compact Objects}

The scope of our discussion has been limited to BHs thus far. In this section, we will try to extend our analysis to the horizonless compact objects, described by the metric in Eq.(\ref{Metric}). However, this time in the absence of a horizon, we need to put the inner boundary condition at the center of the object $r\,=\,0$ as given in Ref.~\cite{Hod:2014ena}. While the behaviors of the auxiliary functions $L(r)$ and $R(r)$ defined earlier remain unchanged at spatial infinity due to asymptotic flatness, the inner boundary conditions at the object's center are as follows: $R(0)\,=\,0$, and $L(0)\,>\,0$, deduced from Ref.~\cite{Hod:2014ena} with the assumption that $h(0) = 0$.\\
 
Following Ref.~\cite{Hod:2014ena}, regularity of the matter configurations requires various components of $T^\mu_\nu$ such as $\rho$, and $p$ to be finite and well-behaved at $r=0$. Thus, we have $\hat{P}(r)\,=\,0$ at the center of the compact object. Now, for the purpose of extending the no-short hair result for horizonless objects, we need to assume that the trace $T$ of the energy-momentum tensor $T^{\mu}_{\nu}$ is non-negative ($T\,\geq\,0$) \cite{Peng:2020hkz}, which is exactly opposite to the BH scenario. While the other two assumptions on the matter, namely the WEC and asymptotic fall-off of $\hat{P}(r)$ remain unchanged as in the case of BHs.\\

Just like the BH scenario,  under these conditions the quantity $\Delta(r)\geq \,0$ in the region $0\,\leq\,r\,\leq\,r_\gamma^1$, where $r_\gamma^1$ is the location of the innermost LR. Then, we must have $\hat{P}'(r)\geq\,0$ in the same region. This statement, along with Eq.(\ref{rconserve}) implies that $\hat{P}(r)$ starts with a zero value at the center of the compact object and then, it increases with $r$ at least up to $r=r_\gamma^1$. If we define $r\,=\,r_{\text{hair}}$ to be the radius at which $|\hat{P}(r)|$ has a local maximum, then we must have $r_{\text{hair}}\,\geq\,r_\gamma^1$.  
Thus, for horizonless compact objects also, the hair is extended at least up to the innermost LR. This generalizes the results of Ref.~\cite{Peng:2020hkz} in a theory-agnostic fashion.\\

Few comments on this result are as follows. First, note that the condition $T > 0$ is not satisfied by ``usual'' matter content, since for those $\rho \gg (p,\, p_T)$. Therefore, for ordinary celestial objects, the no-short hair theorem is not applicable. However, for objects made of `exotic' matter, $T>0$ condition may hold true and then, the no-short hair result can give us useful information about their structures. Another point to note is that, the asymptotic fall-off condition of $\hat{P}(r)$ makes sure of the existence of LRs (an even number of them); otherwise in the absence of LRs, $\hat{P}(r)$ would increase monotonically with $r$.

\section{Conclusion and Discussions}
In this work, we have genralized the no-short hair theorem of GR discussed in Refs.\cite{Nunez:1996xv, Hod:2011aa} in a theory-agnostic way, which is valid for arbitrary spacetime dimensions ($D \geq 4$). Such a result has both theoretical and observational significance. On theory side, it helps us understand some key features of BH solutions of modified gravity in a unified way. Additionally, this  theorem have relevance for both EHT (shadow) and GW (ringdown quasi-normal modes) observations that could probe the near-field region of a BH. Since the the hairosphere cannot be confined solely in the near-horizon regime, exploring the near-LR region alone might give us important information about the signatures of BH hairs. Moreover, the presence of hairs near the LR may give rise to intriguing new phenomena, including gravitational lensing effects and potential modifications to the BH shadow.\\

Apart from these interesting results, we have discussed several useful consequences of our analysis. For example, we have shown that it is possible to set an upper bound on the size of innermost LR of various theories of gravity including EGB theory in $D \geq 5$ dimensions. This bound may be translated to constraint both the shadow size \cite{Chakraborty:2021dmu}, and the real part of the eikonal quasi-normal modes of EGB BHs under perturbation \cite{Hod:2013jhd}. Also, a few comments on the possible generalization of the no-short hair theorem for horizonless compact objects are outlined.\\

The implications of our findings are significant and warrant further consideration. For instance, it will be interesting to extend our work when various assumptions, such as spherical symmetry, asymptotic flatness or WEC on matter are relaxed. Especially, it will be of great observational relevance if the short-hair behaviour of rotating BHs could be investigated \cite{Hod:2014sha}. We leave these for a future attempt.

\section{Acknowledgement}
We thank Shahar Hod for discussion and helpful comments. The research of R.G. is supported by the Prime Minister Research Fellowship (PMRF ID: 1700531), Government of India. SK acknowledges the support from the Sabarmati Bridge Fellowship (Project ID: MIS/IITGN/R$\&$D/SS/202223/047) from IIT Gandhinagar. S.~S. acknowledges support from the Department of Science and Technology, Government of India under the SERB CRG Grant (CRG/2020/004562).

\end{document}